\documentclass[11pt]{article}
\usepackage[latin9]{inputenc}
\usepackage{amsmath}
\usepackage{amssymb}
\usepackage{esint}
\usepackage[unicode=true,
 bookmarks=false,
 breaklinks=false,pdfborder={0 0 1},backref=section,colorlinks=false]
 {hyperref}
\usepackage{breakurl}

\makeatletter
\@ifundefined{date}{}{\date{}}

\usepackage{esint}
\usepackage{latexsym}
\usepackage{graphicx}\usepackage{bm}\usepackage{longtable}

\usepackage{xcolor}

\@addtoreset{equation}{section}

\makeatother

\begin{document}

\title{Glueball-baryon interactions in holographic QCD}

\maketitle
\begin{center}
Si-wen Li\footnote{Email: cloudk@mail.ustc.edu.cn} 
\par\end{center}

\begin{center}
\emph{Department of Modern Physics, }\\
\emph{ University of Science and Technology of China, }\\
\emph{ Hefei 230026, Anhui, China}\\
\emph{ }
\par\end{center}

\vspace{10mm}

\begin{abstract}
Studying the Witten-Sakai-Sugimoto model with type IIA string theory,
we find the glueball-baryon interaction is predicted in this model.
The glueball is identified as the 11D gravitational waves or graviton
described by the M5-brane supergravity solution. Employing the relation
of M-theory and type IIA string theory, glueball is also 10D gravitational
perturbations which are the excited modes by close strings in the
bulk of this model. On the other hand, baryon is identified as a D4-brane
wrapped on $S^{4}$ which is named as baryon vertex, so the glueball-baryon
interaction is nothing but the close string/baryon vertex interaction
in this model. Since the baryon vertex could be equivalently treated
as the instanton configurations on the flavor brane, we identify the
glueball-baryon interaction as ``graviton-instanton'' interaction
in order to describe it quantitatively by the quantum mechanical system
for the collective modes of baryons. So the effective Hamiltonian
can be obtained by considering the gravitational perturbations in
the flavor brane action. With this Hamiltonian, the amplitudes and
the selection rules of the glueball-baryon interaction can be analytically
calculated in the strong coupling limit. We show our calculations
explicitly in two characteristic situations which are ``scalar and
tensor glueball interacting with baryons''. Although there is a long
way to go, our work provides a holographic way to understand the interactions
of baryons in hadronic physics and nuclear physics by the underlying
string theory.
\end{abstract}
\newpage{}

\section{Introduction}

The underlying fundamental theory QCD for nuclear physics and particle
physics has achieved great successes. However, nuclear physics remains
one of the most difficult and intriguing branches of high energy physics
because physicists are still unable to analytically predict the behavior
of nuclei or even a single proton. The key problem is that the behavior
in the strong-coupling regime of QCD is less clear theoretically.
Fortunately, gauge/gravity (gauge/string) duality (See, e.g., \cite{key-001,key-002,key-003,key-004,key-005}
for a review) has become a revolutionary and powerful tool for studying
the strongly coupled quantum field theory. Particularly, the Witten-Sakai-Sugimoto
(WSS) model \cite{key-006,key-007,key-008}, as one of the most famous
models, has been proposed to holographically study the non-perturbative
QCD for a long time \cite{key-009,key-010,key-011,key-012,key-013,key-014,key-015,key-016,key-017,key-018,key-019}.
Therefore, in this paper, we are going to extend the previous works
to study the interactions in holographic QCD.

The holographic glueball-meson interaction has been studied in \cite{key-020,key-021,key-022,key-023}
by naturally considering the gravitational waves or graviton in the
bulk of this model. Since the gravitational waves or graviton signals
the glueball states holographically and mesons are excited by the
open string on the flavor branes, the close/open string (on the flavor
brane) interaction is definitely interpreted as glueball-meson interaction.
And the effective action could be derived by taking account of the
gravitational perturbation in the flavor brane action.

On the other hand, in the WSS model, baryon could be identified as
a $\mathrm{D}4^{\prime}$-brane\footnote{In order to distinguish from $N_{c}$ D4-branes who are responsible
for the background geometry, we denote the baryon vertex as ``$\mathrm{D}4^{\prime}$-brane''
in this paper.} wrapped on $S^{4}$, which is named as ``baryon vertex'' \cite{key-024,key-025}.
The $\mathrm{D}4^{\prime}$-brane has to attach the ends of $N_{c}$
fundamental strings since the $S^{4}$ is supported by $N_{c}$ units
of a R-R flux in the supergravity (SUGRA) solution. Such a $\mathrm{D}4^{\prime}$-brane
is realized as a small instanton configuration in the world-volume
theory of the flavor branes in this model. Basically, the baryon states
could be obtained by quantizing the baryon vertex. In the strong coupling
limit (i.e. the t'Hooft coupling constant $\lambda\gg1$), the two-flavor
case (i.e. $N_{f}=2$) has been studied in \cite{key-009} and it
turns out that baryons can be described by the $SU(2)$ Belavin-Polyakov-Schwarz-Tyupkin
(BPST) instanton solution with a $U\left(1\right)$ potential in the
world-volume theory of the flavor branes. And employing the soliton
picture, baryon states could be obtained by a holographic quantum
mechanical system for collective modes, see also Appendix B.

Accordingly, there must be close string/$\mathrm{D}4^{\prime}$-brane
interaction if the baryon vertex is taken into account, which could
be interpreted as the glueball-baryon interaction in this model. Thus
we will explore whether or not it is able to describe this interaction
by the quantum mechanical system (\ref{eq:43}). So the main contents
of this paper are: First, we find that there must be the glueball-baryon
interaction in this holographic model. Second, we use the holographic
quantum mechanical system in \cite{key-009} (or Appendix B) to describe
the the glueball-baryon interaction quantitatively in the $\lambda\gg1$
limit. Since the analytical instanton configuration with generic numbers
of the flavors is not known, only the two-flavor case (i.e. $N_{f}=2$)
\cite{key-009} is considered in this paper. The outline of this paper
is very simple. In Section 2, we discuss the glueball-baryon interaction
in this model and how to describe it quantitatively in the strong
coupling limit. Section 3 is the Summary and discussion. Since there
are many papers and lectures about the WSS model (such as \cite{key-006,key-014}),
we will not review this model systematically. Only the relevant parts
of this model are collected in Appendix A and B on which our discussions
and calculations are based. Appendix C shows some details of the calculation
in our manuscript.

\section{The equivalent description of the glueball-baryon interactions}

In this section, we will explore that how the ``close string interacting
with baryon vertex'' can be interpreted as the ``glueball interacting
with baryon'' and how to describe it by the holographic quantum mechanical
system in \cite{key-009} (or in Appendix B). First of all, let us
take a look at the most general aspects about the interaction of the
graviton (close string) in this model. 

As a gravity theory, it is very natural to consider the gravitational
waves (or graviton) in the bulk of this D4/D8 system (WSS model).
According to \cite{key-020,key-021,key-022,key-023}, such a gravitational
perturbation signals the glueball states and can definitely interact
with the open string on the flavor branes. Thus such close/open string
interactions have been holographically interpreted as ``glueball-meson
interactions'' or ``glueball decays to the mesons'' because mesons
are excited by the open strings on the flavor branes in this model.
Interestingly, once the baryon vertex is taken into account, the interaction
between baryon vertex and graviton (or gravitational waves) must occur
since the graviton is excited by the close string in the bulk. Hence
there must be close string/baryon vertex interaction which could be
interpreted as glueball-baryon interaction in this model. It provides
a holographic way for understanding and can be treated as a parallel
mechanism to ``glueball-meson interaction'' proposed in \cite{key-020,key-021,key-022,key-023}.

Basically, the ``glueball-baryon interaction'' in this model is
nothing but the close/open string (D-brane, baryon vertex) interaction.
However, it is not easy to quantitatively describe the close/open
string or close string/D-brane interaction by the underlying string
theory in a generic spacetime, in order to describe glueball interacting
with baryons. Fortunately, according to \cite{key-006,key-009,key-010,key-012,key-013,key-014,key-024,key-025},
baryon vertex is equivalently described by the instanton configuration
in world-volume of the $\mathrm{D}8/\overline{\mathrm{D}8}$-branes
with the BPST solution (\ref{eq:38}). Therefore, the ``glueball-baryon
interaction'' could be identified as the ``gravitons (or gravitational
waves) interacting with instantons'' in the world-volume theory of
the flavor branes.

With this idea, let us consider a gravitational perturbation in the
bulk geometry since the glueball states are signaled by the graviton
in this model, i.e. replace the metric as,

\begin{equation}
g_{MN}\rightarrow g_{MN}^{\left(0\right)}+h_{MN},\label{eq:1}
\end{equation}
where $g_{MN}^{(0)}$ is the background metric (\ref{eq:28}) and
$h_{MN}$ is a perturbative tensor which satisfies $h_{MN}\ll g_{MN}^{(0)}$.
Then, we consider the interaction (coupling) between graviton (glueball)
and instantons in the world-volume of the flavor branes. The dynamic
in the world-volume theory of the flavor branes is described by the
Yang-Mills action (\ref{eq:32}) plus the Chern-Simons (CS) action
(\ref{eq:35}). Since the CS action (\ref{eq:35}) is independent
of the metric, it remains to be (\ref{eq:35}) even if (\ref{eq:1})
is imposed. However the Yang-Mills action (\ref{eq:32}) contains
additional terms which depend on $h_{MN}$ as,

\begin{align}
S_{YM} & =S_{YM}^{(0)}+S_{YM}^{(1)}+\mathcal{O}\left(h_{MN}^{2}\right),\nonumber \\
S_{YM}^{(0)} & =-\frac{1}{4}(2\pi\alpha')^{2}T_{8}\int_{\mathrm{D}8/\overline{\mathrm{D}8}}d^{4}xdUd\Omega_{4}e^{-\Phi^{(0)}}\sqrt{-\det g_{ab}^{(0)}}\mathrm{Tr}\left[g^{(0)ac}g^{(0)bd}\mathcal{F}_{ab}\mathcal{F}_{cd}\right],\nonumber \\
S_{YM}^{(1)} & =\frac{1}{4}(2\pi\alpha')^{2}T_{8}\int_{\mathrm{D}8/\overline{\mathrm{D}8}}d^{4}xdUd\Omega_{4}e^{-\Phi^{(0)}}\left(1-\delta\Phi\right)\sqrt{-\det g_{ab}^{(0)}}\times\nonumber \\
 & \ \ \ \ \mathrm{Tr}\left[\left(h^{ac}g^{(0)bd}+h^{bd}g^{(0)ac}\right)\mathcal{F}_{ab}\mathcal{F}_{cd}\right]\label{eq:2}
\end{align}
Notice that only the linear perturbation of $h_{MN}$ is considered
in (\ref{eq:2}). So if $h_{MN}$ is the mode of gravitational waves
propagating in the bulk, it must depend on time which means $h_{MN}=h_{MN}\left(t\right)$.
Furthermore, because the baryon states are given by the quantum mechanical
system (Appendix B), so once we evaluate the potential term (\ref{eq:43})
in the moduli space by using (\ref{eq:42}) with (\ref{eq:2}), it
implies there must be an additionally time-dependent term $H\left(t\right)$
to the Hamiltonian (\ref{eq:43}). Therefore, the transition amplitude
can be calculated by the standard technique of time-dependent perturbation
in quantum mechanics in order to describe the glueball-baryon interaction
quantitatively. So let us evaluate the perturbed Hamiltonian $H\left(t\right)$
and the transition amplitude explicitly by taking account of two characteristic
situations in the following subsections.

\subsection{Interactions with scalar glueball}

In this section, let us consider the ``scalar glueball interacting
with baryons''. In order to evaluate the perturbed Hamiltonian, we
need to write the explicit formulas of the gravitational perturbation
first. In this model, the scalar glueball can be described by the
gravitational polarization \cite{key-021,key-022,key-023}. So let
us consider the gravitational polarization in 11D SUGRA because the
WSS model is based on type IIA SUGRA which can be reduced from M5-brane
solution of 11D SUGRA for M-theory (See \cite{key-026,key-027} for
a complete review). For scalar glueball, the gravitational polarization
in 11D SUGRA takes the following forms,

\begin{eqnarray}
H_{44} & = & -\frac{r^{2}}{L^{2}}F\left(r\right)H_{E}\left(r\right)G_{E}\left(x\right),\nonumber \\
H_{\mu\nu} & = & \frac{r^{2}}{L^{2}}H_{E}\left(r\right)\left[\frac{1}{4}\eta_{\mu\nu}-\left(\frac{1}{4}+\frac{3r_{KK}^{6}}{5r^{6}-2r_{KK}^{6}}\right)\frac{\partial_{\mu}\partial_{\nu}}{M_{E}^{2}}\right]G_{E}\left(x\right),\nonumber \\
H_{55} & = & \frac{r^{2}}{4L^{2}}H_{E}\left(r\right)G_{E}\left(x\right),\nonumber \\
H_{rr} & = & -\frac{L^{2}}{r^{2}}\frac{1}{F\left(r\right)}\frac{3r_{KK}^{6}}{5r^{6}-2r_{KK}^{6}}H_{E}\left(r\right)G_{E}\left(x\right),\nonumber \\
H_{r\mu} & = & \frac{90r^{7}r_{KK}^{6}}{M_{E}^{2}L^{2}\left(5r^{6}-2r_{KK}^{6}\right)^{2}}H_{E}\left(r\right)\partial_{\mu}G_{E}\left(x\right),\label{eq:3}
\end{eqnarray}
We use $G_{AB},\ H_{AB}$ to represent the 11D metric and gravitational
polarizations in oder to distinguish 10D metric $g_{MN}$ and perturbation
$h_{MN}$. The 11 coordinates correspond to $\left\{ x^{\mu},x^{4},x^{5},r,\Omega_{4}\right\} $
and $x=\left\{ x^{\mu}\right\} $ in our convention. The function
$F\left(r\right)$ and the relation between 11D ($r,\ r_{KK},\ L$)
and 10D variables ($U,\ z,\ U_{KK},\ R$) are given as,

\begin{equation}
F\left(r\right)=1-\frac{r_{KK}^{6}}{r^{6}},\ U=\frac{r^{2}}{2L},\ 1+\frac{z^{2}}{U_{KK}^{2}}=\frac{r^{6}}{r_{KK}^{6}}=\frac{U^{3}}{U_{KK}^{3}},\ L=2R.
\end{equation}
Since the near-horizon solution of M5-branes in 11D is $AdS_{7}\times S^{4}$,
the 11D metric satisfies the equations of motion from the following
action with the integration on $S^{4}$,

\begin{equation}
S_{11\mathrm{D}}=\frac{1}{2\kappa_{11}^{2}}\left(\frac{L}{2}\right)^{4}V_{4}\int d^{7}x\sqrt{-\det G}\left(\mathcal{R}_{11\mathrm{D}}+\frac{30}{L^{2}}\right).\label{eq:5}
\end{equation}
Imposing the near-horizon solution of M5-branes with (\ref{eq:3})
to (\ref{eq:5}), we obtain the eigenvalue equation for $H_{E}\left(r\right)$
as,

\begin{equation}
\frac{1}{r^{3}}\frac{d}{dr}\left[r\left(r^{6}-r_{KK}^{6}\right)\frac{d}{dr}H_{E}\left(r\right)\right]+\left[\frac{432r^{2}r_{KK}^{12}}{\left(5r^{6}-2r_{KK}^{6}\right)^{2}}+L^{4}M_{E}^{2}\right]H_{E}\left(r\right)=0,\label{eq:6}
\end{equation}
and the kinetic action of the function $G_{E}\left(x\right)$ which
is,

\begin{equation}
S_{G_{E}\left(x\right)}=\mathcal{C}_{E}\int d^{4}xdx^{4}dx^{5}\frac{1}{2}\left[\left(\partial_{\mu}G_{E}\right)^{2}+M_{E}^{2}G_{E}^{2}\right],\label{eq:7}
\end{equation}
with

\begin{equation}
\mathcal{C}_{E}=\int_{r_{KK}}^{\infty}dr\frac{r^{3}}{L^{3}}\frac{5}{8}H_{E}^{2}\left(r\right).
\end{equation}
Obviously, (\ref{eq:7}) shows why (\ref{eq:3}) signals scalar glueball
field. Then we have to translate 11D gravitational polarization (\ref{eq:3})
into 10D WSS model in order to evaluate the Hamiltonian for collective
modes. Employing the dimensional reduction as \cite{key-026,key-027},
the components of 10D $h_{MN}$ are collected by subtracting $g_{MN}^{(0)}$.
As a result, they are,

\begin{align}
h_{\mu\nu} & =\left(\frac{U}{R}\right)^{3/2}\left[\frac{R}{2U}H_{55}\eta_{\mu\nu}+\frac{R}{U}H_{\mu\nu}\right],\nonumber \\
h_{44} & =\left(\frac{U}{R}\right)^{1/2}\left[H_{44}+\frac{1}{2}f\left(U\right)H_{55}\right],\nonumber \\
h_{zz} & =\frac{4R^{3/2}U_{KK}}{9U^{5/2}}\left(\frac{R}{2U}H_{55}+\frac{U_{KK}z^{2}}{RU^{2}}H_{rr}\right),\nonumber \\
h_{z\mu} & =\frac{2U_{KK}z}{3U^{2}}H_{r\mu},\ \ h_{\Omega\Omega}=\frac{R^{5/2}}{2U^{1/2}}H_{55},\label{eq:9}
\end{align}
with the dilaton,

\begin{equation}
e^{4\Phi/3}=\frac{U}{R}\left(1+\frac{R}{U}H_{55}\right).
\end{equation}
For the reader convenience, we give the explicit form of the equation
(\ref{eq:6}) in the $z$ coordinate, which is,

\begin{align}
0= & \ H_{E}^{\prime\prime}\left(z\right)+\frac{U_{KK}^{2}+3z^{2}}{z\left(U_{KK}^{2}+z^{2}\right)}H_{E}^{\prime}\left(z\right)\nonumber \\
 & +\frac{432U_{KK}^{13/3}\left(U_{KK}^{2}+z^{2}\right)^{1/3}+4R^{3}M_{E}^{2}\left(3U_{KK}^{2}+5z^{2}\right)^{2}}{9U_{KK}^{1/3}(U_{KK}^{2}+z^{2})^{4/3}\left(3U_{KK}^{2}+5z^{2}\right)^{2}}H_{E}\left(z\right).\label{eq:11}
\end{align}
While (\ref{eq:11}) is difficult to solve, we have to search for
a solution for $H_{E}$ in order to evaluate the perturbed Hamiltonian
for collective modes. Sine only the $\mathcal{O}\left(\lambda^{0}\right)$
of the Hamiltonian (\ref{eq:41}) (\ref{eq:43}) is the concern in
our paper, we need to solve (\ref{eq:11}) up to $\mathcal{O}\left(\lambda^{-1}\right)$.
Rescale (\ref{eq:11}) as (\ref{eq:37}), we obtain the following
equation\footnote{``$\boldsymbol{\mathrm{z}}$'' is the rescaled coordinate defined
as in (\ref{eq:37}).} (derivatives are w.r.t. $\boldsymbol{\mathrm{z}}$),

\begin{equation}
H_{E}^{\prime\prime}\left(\boldsymbol{\mathrm{z}}\right)+\left(\frac{1}{\boldsymbol{\mathrm{z}}}+\frac{2\boldsymbol{\mathrm{z}}}{\lambda}\right)H_{E}^{\prime}\left(\boldsymbol{\mathrm{z}}\right)+\frac{16+3M_{E}^{2}}{3\lambda}H_{E}\left(\boldsymbol{\mathrm{z}}\right)+\mathcal{O}\left(\lambda^{-2}\right)=0.\label{eq:12}
\end{equation}
In order to compare our calculations with \cite{key-009}, we have
employed the unit of $M_{KK}=U_{KK}=1$ so that $R^{3}=9/4$. The
(\ref{eq:12}) is easily to solve in terms of hypergeometric function
and Meijer G function which is,

\begin{align}
H_{E}\left(\boldsymbol{\mathrm{z}}\right) & =C_{1}\mathrm{Hypergeometri}\mathrm{c}_{1}\mathrm{F}_{1}\left[\frac{4}{3}+\frac{M_{E}^{2}}{4},1,-\frac{\boldsymbol{\mathrm{z}}^{2}}{\lambda}\right]\nonumber \\
 & +C_{2}\mathrm{MeijerG}\left[\left\{ \#\right\} ,-\frac{1}{3}-\frac{M_{E}^{4}}{4},\left\{ 0,0,\left\{ \#\right\} \right\} ,\frac{\boldsymbol{\mathrm{z}}^{2}}{\lambda}\right],\label{eq:13}
\end{align}
where $C_{1},C_{2}$ are two integration constants. Since the background
is the bubble solution of D4-branes, the metric in our model must
be regular everywhere. Accordingly, we have to set $C_{2}=0$ because
Meijer G function diverges at $U=U_{KK}=1$. On the other hand, $C_{1}$
has to consistently satisfy $C_{1}\ll1$, because $H_{E}$ appearing
in (\ref{eq:9}) should also be the perturbation to the background
metric $g_{MN}^{(0)}$. Therefore the solution of $H_{E}$ is only
valid up to $\mathcal{O}\left(\lambda^{-1}\right)$ according to (\ref{eq:12}).
Hence, we have the solution of $H_{E}$ in the large $\lambda$ expansion,

\begin{equation}
H_{E}\left(\boldsymbol{\mathrm{z}}\right)\simeq C_{1}-C_{1}\frac{\left(16+3M_{E}^{2}\right)\boldsymbol{\mathrm{z}}^{2}}{12\lambda}+\mathcal{O}\left(\lambda^{-2}\right).\label{eq:14}
\end{equation}

Next, we will evaluate the perturbed Hamiltonian with (\ref{eq:9})
additional to (\ref{eq:43}). Using (\ref{eq:42}) and (\ref{eq:2}),

\begin{equation}
S_{YM}^{(0)}+S_{YM}^{(1)}+S_{CS}=-\int dt\left[U^{(0)}(X^{\alpha})-H\left(t,X^{\alpha}\right)\right].\label{eq:15}
\end{equation}
$U^{(0)}(X^{\alpha})$ is obtained by evaluating $S_{YM}^{(0)}+S_{CS}$
which is the exact forms of the potential in (\ref{eq:43}). Therefore
we need to evaluate $S_{YM}^{(1)}$ in order to obtain the perturbed
Hamiltonian $H\left(t,X^{\alpha}\right)$ in (\ref{eq:15}) and the
procedures are as follows,
\begin{enumerate}
\item We decompose the $U\left(2\right)$ gauge field in $S_{YM}^{(1)}$
(\ref{eq:2}) as (\ref{eq:36}) and use (\ref{eq:38}) to represent
the instanton (baryon) in the world-volume of $\mathrm{D}8/\overline{\mathrm{D}8}$-branes.
\item Insert (\ref{eq:9}) into $S_{YM}^{(1)}$ (\ref{eq:2}), rescale the
obtained formula of $S_{YM}^{(1)}$ by imposing (\ref{eq:37}) and
then expand the result up to $\mathcal{O}\left(\lambda^{-1}\right)$. 
\item Finally, we use (\ref{eq:15}) to evaluate the perturbed Hamiltonian
for the collective modes up to $\mathcal{O}\left(\lambda^{-1}\right)$.
\end{enumerate}
\noindent While the above procedures are quite straightforward, the
calculation is very messy. So let us give the resultant formula here\footnote{$\kappa$ is given in (\ref{eq:44}) and more details of the calculation
for (\ref{eq:17}) are given in the Appendix C. }. The perturbed Hamiltonian can be written as,

\begin{equation}
H_{Scalar}\left(t,X^{\alpha}\right)=\mathcal{A}\kappa\int d^{3}\boldsymbol{\mathrm{x}}d\boldsymbol{\mathrm{z}}d\Omega_{4}\mathcal{K}\left(t,\boldsymbol{\mathrm{x}},\boldsymbol{\mathrm{z}},X^{\alpha}\right),
\end{equation}
where the function $\mathcal{K}\left(t,\boldsymbol{\mathrm{x}},\boldsymbol{\mathrm{z}},X^{\alpha}\right)$
is given in (\ref{eq:50}) in the unit of $U_{KK}=M_{KK}=1$. Hence
the Hamiltonian is calculated as,

\begin{align}
H_{Scalar}\left(t,X^{\alpha}\right)= & \mathcal{A}\kappa C_{1}\cos\left(\omega t\right)\bigg\{\frac{9}{2}\pi^{2}\left(2-\frac{5\omega^{2}}{M_{E}^{2}}\right)+\nonumber \\
 & +\bigg[\left(\frac{40k^{2}-18M_{E}^{4}+420\omega^{2}}{16M_{E}^{2}}+\frac{48+45\omega^{2}}{16}\right)\left(2Z^{2}+\rho^{2}\right)\pi^{2}\nonumber \\
 & +\frac{9M_{E}^{2}-90\omega^{2}-45k^{2}}{1280M_{E}^{2}a^{2}\pi^{2}\rho^{2}}+9k^{2}\rho^{2}\pi^{2}\left(\frac{5\omega^{2}}{8M_{E}^{2}}-\frac{1}{4}\right)\bigg]\lambda^{-1}+\mathcal{O}\left(\lambda^{-2}\right)\bigg\}.\label{eq:17}
\end{align}
$\mathcal{A}$ is a constant independent on $\lambda$\footnote{In the unit of $M_{KK}=U_{KK}=1$, $\mathcal{A}$ should be $\mathcal{A}=\frac{243}{512\pi^{2}l_{s}^{2}}$.}.
$k$ and $\omega$ is the 3-momentum and the frequency of the glueball
field $G_{E}\left(t,\boldsymbol{\mathrm{x}}\right)$\footnote{Since our theory is symmetrically rotated in the 3d $x^{i}$-space,
we assume the momentum $k$ of the glueball field $G\left(t,\boldsymbol{\mathrm{x}}\right)$
is along $x^{3}$ direction. }. Notice that (\ref{eq:17}) is suitable to be a perturbation since
$C_{1}$ has to satisfy $C_{1}\ll1$. Hence, with the quantum mechanical
system of baryons (\ref{eq:43}), the average transition amplitude
$\mathcal{M}$ and the probability of transition $\mathcal{P}$ can
be evaluated by the standard technique in the quantum mechanics with
a time-dependent perturbation, which is,

\begin{align}
\mathcal{P}_{i\to f} & =\left|\int_{0}^{t}\langle H\left(t^{\prime},X^{\alpha}\right)e^{-iE_{if}t^{\prime}}\rangle dt^{\prime}\right|^{2},\nonumber \\
 & =\left|\int_{0}^{t}e^{i\left(E_{if}-\omega\right)t^{\prime}}\mathcal{M}\left(i\rightarrow j\right)dt^{\prime}\right|,\label{eq:18}
\end{align}
where $E_{ij}=E\left(l^{\prime},n_{\rho}^{\prime},n_{z}^{\prime}\right)-E\left(l,n_{\rho},n_{z}\right)$
is defined by (\ref{eq:45}). For simplification, let us consider
the case of small $k$ limit i.e. $k\rightarrow0$ which means the
glueball field, as an external field, is homogeneous. In this limit,
it implies $\omega\simeq M_{E}$ since the ``classical glueball field
$G_{E}\left(t,\boldsymbol{\mathrm{x}}\right)$'' means the onshell
condition $\omega^{2}-k^{2}=M_{E}^{2}$ has to be satisfied. Thus
in small $k$ limit, we can simplify (\ref{eq:17}) as,

\begin{align}
H_{Scalar}\left(t,X^{\alpha}\right)= & \mathcal{A}\kappa C_{1}\cos\left(\omega t\right)\bigg\{-\frac{27}{2}\pi^{2}+\left[-\frac{81}{1280a^{2}\pi^{2}\rho^{2}}+\frac{27M_{E}^{2}+468}{16}\pi^{2}\left(2Z^{2}+\rho^{2}\right)\right]\lambda^{-1}\nonumber \\
 & +\mathcal{O}\left(\lambda^{-2}\right)\bigg\}.\label{eq:19}
\end{align}
By analyzing the eigenfunctions (\ref{eq:45}) of the Hamiltonian
(\ref{eq:43}), we find the following selection rules,

\begin{equation}
\bigg\{\begin{array}{c}
\tilde{l}^{\prime}=\tilde{l}\ \left(l^{\prime}=l\right)\\
n_{z}^{\prime}=n_{z}
\end{array}\ \mathrm{or}\ \bigg\{\begin{array}{c}
\tilde{l}^{\prime}=\tilde{l}\ \left(l^{\prime}=l\right)\\
n_{z}^{\prime}=n_{z}\pm2\\
n_{\rho}^{\prime}=n_{\rho}
\end{array}.\label{eq:20}
\end{equation}
Working out (\ref{eq:18}), it is easy to find another constraint
of the transition which is $\omega=E_{ij}$. Interestingly, our holographical
quantum mechanical system is very similar as the atomic spectrum of
hydrogen. The ``holographic baryon interacting with glueball'' behaves
similarly as the ``electron interacting with photon'' in the hydrogen
atomic. Both of them can be described by the quantum mechanics which
means the baryon (electron) is described by the quantum mechanics
while the glueball (photon), as a classically external field, is described
by the classical gravity theory (classical electrodynamics) respectively.

Furthermore, let us examine whether or not the transition procedures,
in this quantum mechanical system with the constraints and selection
rule discussed above, are really possible to occur. We consider the
low energy (small momentum) limit as the most simple case which is
$k\rightarrow0$, so that $\omega\simeq M_{E}$. Since $M_{E}$ represents
the mass spectrum of the scalar glueball in (\ref{eq:11}), it reads
with the WKB approximation \cite{key-020} ($\bar{\beta}=2\pi$ in
the unit of $M_{KK}=U_{KK}=1$),

\begin{equation}
M_{E}\left(j\right)\simeq\frac{8.12}{\bar{\beta}}\sqrt{j\left(j+\frac{5}{2}\right)}.
\end{equation}
Consequently, we find the following transitions,

\begin{equation}
\frac{E\left(l=1,3;\ n_{\rho}=3;\ n_{z}=0,1,2,3\right)-E\left(l=1,3;\ n_{\rho}=0;\ n_{z}=0,1,2,3\right)}{M_{E}\left(j=1\right)}\simeq1.013,\label{eq:22}
\end{equation}
and 

\begin{equation}
\frac{E\left(l=1,3;\ n_{\rho}=5;\ n_{z}=0,1,2,3\right)-E\left(l=1,3;\ n_{\rho}=0;\ n_{z}=0,1,2,3\right)}{M_{E}\left(j=2\right)}\simeq1.053,\label{eq:23}
\end{equation}
are possible to occur according to the above selection rules and constraint.
Notice that the WSS model is a low-energy effective theory for baryons
or mesons, so it may not be very consistent to consider the high energy
states of baryons in this model. Using (\ref{eq:17}) and (\ref{eq:18}),
the transition ampitude corresponding to (\ref{eq:22}) (\ref{eq:23})
can be calculated, respectively, as,

\begin{align}
\mathcal{M}\left(n_{\rho}=3\rightarrow n_{\rho}^{\prime}=0\right)\bigg|_{l=l^{\prime},n_{z}=n_{z}^{\prime}} & =\frac{\left(27M_{E}^{2}-270\omega^{2}-135k^{2}\right)\mathcal{A}\kappa C_{1}}{44800M_{E}^{2}a^{2}\pi^{2}m_{y}^{4}\omega_{\rho}^{4}\lambda}\nonumber \\
 & \simeq-\frac{243\mathcal{A}\kappa C_{1}}{44800a^{2}\pi^{2}m_{y}^{4}\omega_{\rho}^{4}\lambda}+\mathcal{O}\left(k^{2}\right),\nonumber \\
\mathcal{M}\left(n_{\rho}=5\rightarrow n_{\rho}^{\prime}=0\right)\bigg|_{l=l^{\prime},n_{z}=n_{z}^{\prime}} & =\frac{\left(9M_{E}^{2}-90\omega^{2}-45k^{2}\right)\mathcal{A}\kappa C_{1}}{53760M_{E}^{2}a^{2}\pi^{2}m_{y}^{4}\omega_{\rho}^{4}\lambda}\nonumber \\
 & \simeq-\frac{81\mathcal{A}\kappa C_{1}}{63760a^{2}\pi^{2}m_{y}^{4}\omega_{\rho}^{4}\lambda}+\mathcal{O}\left(k^{2}\right).
\end{align}

\subsection{Interactions with tensor glueball}

Let us consider another special example for the interaction with tensor
glueball. In the bulk, the 11D gravitational polarization of the tensor
glueball could be simply chosen as \cite{key-023},

\begin{equation}
H_{11}=-H_{22}=-\frac{r^{2}}{L^{2}}H_{T}\left(r\right)G_{T}\left(x\right),\label{eq:25}
\end{equation}
where the equation of motion for the radial function $H_{T}$ is,

\begin{equation}
\frac{1}{r^{3}}\frac{d}{dr}\left[r\left(r^{6}-r_{KK}^{6}\right)\frac{d}{dr}H_{T}\left(r\right)\right]+L^{4}M_{T}^{2}H_{T}\left(r\right)=0.
\end{equation}
While (\ref{eq:25}) has to be reduced into 10D metric, it satisfies
the traceless condition,

\begin{equation}
h_{11}=-h_{22}.\label{eq:27}
\end{equation}
Inserting (\ref{eq:27}) into (\ref{eq:2}), we can immediately find
that the perturbed Hamiltonian of the collective coordinates is vanished.
However, the perturbed Hamiltonian from the tensor glueball should
be $H_{Tensor}\left(t,X^{\alpha}\right)\sim\mathcal{O}\left(H_{AB}^{2}\right)$
. Since the gravitational polarization (\ref{eq:25}) is solved by
the linear gravity perturbation, it would be inconsistent to consider
the contribution from $\mathcal{O}\left(H_{AB}^{2}\right)$ to the
Hamiltonian of the collective coordinates.

\section{Summary and discussion}

In this paper, we consider the linearly gravitational perturbation
in the bulk of the Witten-Sakai-Sugimoto model. According to \cite{key-020,key-021,key-022,key-023},
such gravitational perturbations signal the glueball states. On the
other hand, baryon can be identified as wrapped D-brane which is named
as the ``baryon vertex'' as \cite{key-024,key-025}. So in the viewpoints
of the string theory, there must be the glueball-baryon interaction
if the baryon vertex is taken into account. Therefore the glueball-baryon
interaction is nothing but the close string/D-brane (baryon vertex)
interaction in this model. Since baryons can be treated as instanton
configurations in the world-volume of the flavor branes, we identify
the glueball-baryon interaction as ``graviton-instanton'' interaction
as an equivalent description. With the BPST instanton configuration,
we find the perturbed Hamiltonian for the collective modes of the
baryons could be evaluated quantitatively in the strong coupling limit.
Hence the amplitude and the selection rules of the glueball-baryon
interaction can be accordingly calculated. In order to quantitatively
clarify our idea, we show our methods in two characteristic situations
which are ``scalar and tensor glueball interacting with baryons''.
Particularly, the perturbed Hamiltonian of ``tensor glueball-baryon''
interaction is vanished in the linearly gravitational perturbation
which means it should be non-linear interaction in the gravity side.

Our work should be an application of strongly Maldacena's conjecture
since we have considered the quantum effect (graviton) in the gravity
side. So our conclusions may also be suitable with finite $N_{c}$
and $\lambda$. Moreover, if combining \cite{key-020,key-021,key-022,key-023}
with our work, it shows the complete glueball-meson-baryon interaction.
Although these holographic approaches are a little different from
traditional theories, they show us an analytical way to study the
strongly coupled interactions in hadronic physics and nuclear physics
by the string theory.

\section*{ACKNOWLEDGMENTS}

I would like to thank Dr. Chao Wu and Prof. Qun Wang for helpful discussions.
I was supported partially by the Major State Basic Research Development
Program in China under the Grant No. 2015CB856902 and the National
Natural Science Foundation of China under the Grant No. 11125524 for
this work.

\section*{Appendix A: The geometry of the Witten-Sakai-Sugimoto model}

In the WSS model, there are $N_{c}$ coincident D4-branes representing
``colors'' of QCD, wrapped on a supersymmetry breaking compact circle.
The background geometry produced by these D4-branes is described by
10-dimensional type IIA supergravity in the near horizon limit. The
metric reads \cite{key-006}\index{Commands!T!tag@\textbackslash{}tag}
,

\begin{equation}
ds^{2}=\left(\frac{U}{R}\right)^{3/2}\left[\eta_{\mu\nu}dx^{\mu}dx^{\nu}+f\left(U\right)\left(dx^{4}\right)^{2}\right]+\left(\frac{R}{U}\right)^{3/2}\left[\frac{dU^{2}}{f\left(U\right)}+U^{2}d\Omega_{4}^{2}\right],\tag{A-1}\label{eq:28}
\end{equation}
which is the bubble geometry of the D4-brane solution. And the dilaton,
Romand-Romand 4-form field, the function $f\left(U\right)$ are given
as,

\begin{equation}
e^{\phi}\equiv e^{\Phi-\Phi_{0}}=g_{s}\left(\frac{U}{R}\right)^{3/4},\ \ F_{4}=dC_{3}=\frac{2\pi N_{c}}{V_{4}}\epsilon_{4},\ \ f\left(U\right)=1-\frac{U_{KK}}{U^{3}},\tag{A-2}\label{eq:29}
\end{equation}
where $x^{\mu},\mu=0,1,2,3$ and $x^{4}$ are the directions which
the D4-branes are extended along. $U$ is the coordinate of the holographic
radius and $U_{KK}$ is the coordinate radius of the bottom of the
bubble. The relation between $R$ and the string coupling $g_{s}$
with string length $l_{s}$ is given as $R^{3}=\pi g_{s}N_{c}l_{s}^{3}$.
Respectively, $d\Omega_{4}^{2}$, $\epsilon_{4}$ and $V_{4}=8\pi^{2}/3$
are the line element, the volume form and the volume of an $S^{4}$
with unit radius. We have used $x^{4}$ to denote the periodic direction
where the D4-branes are wrapped on as $x^{4}\sim x^{4}+\delta x^{4}$
with $\delta x^{4}=\frac{4\pi}{3}R^{3/2}/U_{KK}^{1/2}$. Accordingly,
the Kaluza-Klein mass can be defined as $M_{KK}=2\pi/\delta x^{4}=\frac{3}{2}U_{KK}^{1/2}/R^{3/2}$.
Hence the parameters $R,\ U_{KK},\ g_{s}$ can be expressed in terms
of QCD variables $g_{YM},\ M_{KK},\ l_{s}$ as,

\begin{equation}
R^{3}=\pi g_{s}N_{c}l_{s}^{3},\ \ U_{KK}=\frac{2}{9}g_{YM}^{2}N_{c}M_{KK}l_{s}^{2},\ \ g_{s}=\frac{1}{2\pi}\frac{g_{YM}^{2}}{M_{KK}l_{s}}.\tag{A-3}
\end{equation}
Additionally, the ``flavors'' of QCD could be introduced into this
model by embedding a stack of $N_{f}$ D8 and anti-D8 branes ($\mathrm{D}8/\overline{\mathrm{D}8}$-branes)
as probes into the background (\ref{eq:28}). The dynamic of the flavor
branes is described by the following action,

\begin{equation}
S_{\mathrm{D}8/\overline{\mathrm{D}8}}=S_{\mathrm{DBI}}+S_{WZ},\tag{A-4}\label{eq:31}
\end{equation}
The first term in (\ref{eq:31}) is the Dirac-Born-Infield (DBI) action
and the second term is the Wess-Zumino (WZ) action. The DBI action
of $\mathrm{D}8/\overline{\mathrm{D}8}$-branes in this model can
be expanded in small field strength. Keeping only $\mathcal{O}(\mathcal{F}^{2})$,
we get the Yang-Mills action for the dual field theory on the flavor
branes, which is\footnote{$\mathcal{F}$ is the dimensionlessful gauge field strength which
is defined as $\mathcal{F}=2\pi\alpha^{\prime}F$.},

\begin{align}
S_{\mathrm{DBI}} & \simeq S_{YM}+\mathcal{O}(\mathcal{F}^{4}),\nonumber \\
S_{YM} & =-\frac{1}{4}(2\pi\alpha')^{2}T_{8}\int_{\mathrm{D}8/\overline{\mathrm{D}8}}d^{4}xdUd\Omega_{4}e^{-\Phi}\sqrt{-\det g_{ab}}\mathrm{Tr}\left[g^{ac}g^{bd}\mathcal{F}_{ab}\mathcal{F}_{cd}\right].\tag{A-5}\label{eq:32}
\end{align}
On the other hand, since only $C_{3}$ in non-vanished (\ref{eq:29}),
the relevant term in WZ action is,

\begin{align}
S_{WZ} & =\frac{1}{3!}\mu_{8}(2\pi\alpha')^{3}\int_{D8}C_{3}\wedge\mathrm{Tr}\mathcal{F}^{3}\nonumber \\
 & =\frac{1}{3!}\mu_{8}(2\pi\alpha')^{3}\int_{D8}dC_{3}\omega_{5}(\mathcal{A}),\tag{A-6}\label{eq:33}
\end{align}
where $\omega_{5}(\mathcal{A})$ is Chern-Simons 5-form given as, 

\begin{equation}
\omega_{5}(\mathcal{A})=\mathrm{Tr}\left(\mathcal{A}\mathcal{F}^{2}-\frac{i}{2}\mathcal{A}^{3}\mathcal{F}-\frac{1}{10}\mathcal{A}^{5}\right).\tag{A-7}
\end{equation}
Since we are going to discuss the two-flavor case i.e. $N_{f}=2$,
the explicit form of (\ref{eq:33}) after integrating out $dC_{3}$
can be written as,

\begin{align}
S_{WZ} & =\frac{N_{c}}{24\pi^{2}}\epsilon_{mnpq}\int d^{4}xdz\bigg[\frac{3}{8}\hat{A}_{0}\mathrm{Tr}(F_{mn}F_{pq})-\frac{3}{2}\hat{A}_{m}\mathrm{Tr}(\partial_{0}A_{n}F_{pq})+\frac{3}{4}\hat{F}_{mn}\mathrm{Tr}(A_{0}F_{pq})\nonumber \\
 & +\frac{1}{16}\hat{A}_{0}\hat{F}_{mn}\hat{F}_{pq}-\frac{1}{4}\hat{A}_{m}\hat{F}_{0n}\hat{F}_{pq}+(\text{total derivatives})\bigg]\equiv S_{CS}.\tag{A-8}\label{eq:35}
\end{align}
where the $U\left(2\right)$ gauge field $\mathcal{A}$ has been decomposed
into its $U(1)$ and $SU\left(2\right)$ part as\footnote{We have used ``$\hat{\ }$'' to represent the Abelian part of the
gauge field while the non-Abelian part is expressed without a ``$\hat{\ }$''.},

\begin{equation}
\mathcal{A}=A^{i}\frac{\tau^{i}}{2}+\frac{1}{\sqrt{2N_{f}}}\hat{A}\times\boldsymbol{1}_{N_{f}\times N_{f}}.\tag{A-9}\label{eq:36}
\end{equation}
Notice that (\ref{eq:35}) is expressed in the $z$ coordinate with
transformation $U^{3}=U_{KK}^{3}+U_{KK}z^{2}$ and the index is defined
as $m,n,p,q=1,2,3,z$ in the above equation. $\tau^{i}$s are the
Pauli matrices. Hence we have used the Yang-Mill action (\ref{eq:32})
plus Chern-Simons action (\ref{eq:35}) to govern the low energy dynamics
on the flavored $\mathrm{D}8/\overline{\mathrm{D}8}$-branes in this
paper.

\section*{Appendix B: Baryon as instanton in the Witten-Sakai-Sugimoto model}

In the WSS model, baryon has been provided by a $\mathrm{D}4^{\prime}$-brane
wrapped on $S^{4}$, which is named as ``baryon vertex''. In the
world-volume theory of the flavor branes, the coordinates $x^{M}$
and the $U\left(2\right)$ gauge field $\mathcal{A}_{M}$ need to
be rescaled as \cite{key-009} in order to obtain the variables independent
of $\lambda$,\index{Commands!T!tag@\textbackslash{}tag},

\begin{align}
x^{m}=\lambda^{-1/2}\boldsymbol{\mathrm{x}}^{m} & ,\ x^{0}=\boldsymbol{\mathrm{x}}^{0},\nonumber \\
\mathcal{A}_{0}\left(t,x\right)=\boldsymbol{\mathcal{A}}_{0}\left(t,x\right) & ,\ \mathcal{A}_{m}\left(t,x\right)=\lambda^{1/2}\boldsymbol{\mathcal{A}}_{m}\left(t,x\right),\nonumber \\
\mathcal{F}_{0m}\left(t,x\right)=\lambda^{1/2}\boldsymbol{\mathcal{F}}_{0m}\left(t,x\right) & ,\ \mathcal{F}_{mn}\left(t,x\right)=\lambda\boldsymbol{\mathcal{F}}_{mn}\left(t,x\right),\tag{B-1}\label{eq:37}
\end{align}
in the expansion of $\lambda^{-1}$. Then by solving the equations
of motion the resultantly non-vanished components of the gauge field
take the following forms,

\begin{align}
\hat{\boldsymbol{\mathrm{A}}}_{0} & =\frac{1}{8\pi^{2}a}\frac{1}{\xi^{2}}\left[1-\frac{\rho^{4}}{\left(\rho^{2}+\xi^{2}\right)^{2}}\right],\nonumber \\
\boldsymbol{\mathrm{F}}_{ij} & =Q\left(\xi,\rho\right)\epsilon_{ijk}\tau^{k},\nonumber \\
\boldsymbol{\mathrm{F}}_{zi} & =Q\left(\xi,\rho\right)\delta_{ij}\tau^{j},\nonumber \\
Q\left(\xi,\rho\right) & =\frac{2\rho^{2}}{\left(\xi^{2}+\rho^{2}\right)^{2}},\ \ a=\frac{1}{216\pi^{3}},\tag{B-2}\label{eq:38}
\end{align}

\noindent where

\begin{equation}
\xi^{2}=\left(\vec{\boldsymbol{\mathrm{x}}}-\vec{X}\right)^{2}+\left(\boldsymbol{\mathrm{z}}-Z\right)^{2}.\tag{B-3}
\end{equation}
In (\ref{eq:38}), we have used same convention as \cite{key-009},
so that $\vec{\boldsymbol{\mathrm{x}}}=\left\{ \boldsymbol{\mathrm{x}}^{i}\right\} ,\ i=1,2,3$
represents the 3-spatial coordinates where the baryons or instantons
live and $\rho$ represents its size. According to \cite{key-009}
(See \cite{key-028} for a complete review), the baryon spectrum could
be obtained by a quantum mechanical system for the collective coordinates
in a moduli space of one instanton. Since we are working in the strong
coupling limit (i.e. $\lambda\gg1$), the contribution of $\mathcal{O}\left(\lambda^{-1}\right)$
could be neglected. Accordingly the moduli space takes the following
topology,

\begin{equation}
\mathcal{M}=\mathbb{R}^{4}\times\mathbb{R}^{4}/\mathbb{Z}_{2}.\tag{B-4}
\end{equation}

\noindent The first $\mathbb{R}^{4}$ corresponds to the position
of the instanton which is parameterized by the collective coordinates
$\left(\overrightarrow{X},Z\right)$ and $\mathbb{R}^{4}/\mathbb{Z}_{2}$
is parameterized by the size $\rho$ and the $SU(2)$ orientation
of the instanton. $\mathbb{R}^{4}/\mathbb{Z}_{2}$ can be parametrized
by $y_{I},\ I=1,2,3,4$ and the size of the instanton corresponds
to the radial coordinate i.e. $\rho=\sqrt{y_{1}^{2}+...y_{4}^{2}}$.
The $SU(2)$ orientation is parameterized by $a_{I}=\frac{y_{I}}{\rho}$
with the normalized constraint $\sum_{I=1}^{4}a_{I}^{2}=1$\footnote{Such a parameterization is also used in \cite{key-011,key-029,key-031,key-030}.}.
It has been turned out that the Lagrangian of the collective coordinates
in such a moduli space is given as,

\begin{equation}
L=\frac{m_{X}}{2}g_{\alpha\beta}\dot{X^{\alpha}}\dot{X^{\beta}}-U\left(X^{\alpha}\right)+\mathcal{O}\left(\lambda^{-1}\right).\tag{B-5}\label{eq:41}
\end{equation}
The first term in (\ref{eq:41}) is the line element of the moduli
space which corresponds the kinetic term in the Lagrangian while the
second term corresponds the potential of this quantum mechanical system.
Notice that we have used $X^{\alpha}=\left(\overrightarrow{X},\ Z,\ y_{I}\right)$,
and $m_{X}=8\pi^{2}aN_{c}$. The potential term $U\left(X^{\alpha}\right)$
in (\ref{eq:41}) could be calculated by employing the soliton picture
as \cite{key-009,key-011,key-028,key-029,key-030,key-031}, which
takes the following form,

\begin{equation}
S_{\mathrm{D}8/\overline{\mathrm{D}8}}^{\mathrm{onshell}}\simeq S_{YM+CS}^{\mathrm{onshell}}=-\int dtU(X^{\alpha}).\tag{B-6}\label{eq:42}
\end{equation}
After quantization, the Hamiltonian corresponding to (\ref{eq:41})
for the collective coordinates is given as,

\noindent 
\begin{align}
H & =M_{0}+H_{y}+H_{Z}+\mathcal{O}\left(\lambda^{-1}\right),\nonumber \\
H_{y} & =-\frac{1}{2m_{y}}\sum_{I=1}^{4}\frac{\partial^{2}}{\partial y_{I}^{2}}+\frac{1}{2}m_{y}\omega_{y}^{2}\rho^{2}+\frac{\mathcal{Q}}{\rho^{2}},\nonumber \\
H_{Z} & =-\frac{1}{2m_{Z}}\frac{\partial^{2}}{\partial Z^{2}}+\frac{1}{2}m_{Z}\omega_{Z}^{2}Z^{2},\tag{B-7}\label{eq:43}
\end{align}
where\footnote{Eqs. (\ref{eq:43}) - (\ref{eq:46}) are expressed in the unit of
$M_{KK}=U_{KK}=1$.},

\begin{equation}
M_{0}=8\pi^{2}\kappa,~\ \omega_{Z}^{2}=\frac{2}{3},~~\omega_{\rho}^{2}=\frac{1}{6},~~\mathcal{Q}=\frac{N_{c}}{40\pi^{2}a},\ \ \kappa=\frac{\lambda N_{c}}{216\pi^{3}}.\tag{B-8}\label{eq:44}
\end{equation}
The eigenfunctions and eigenvalues of (\ref{eq:43}) can be easily
evaluated by solving its Schrodinger equation, respectively they are\footnote{The relation of $l$ and $\tilde{l}$ is $\tilde{l}=-1+\sqrt{\left(l+1\right)^{2}+2m_{y}\mathcal{Q}}$
and the quantum number of the angle momentum can be represented by
either $l$ or $\tilde{l}$.},

\begin{align}
\psi(y_{I}) & =R(\rho)T^{(l)}(a_{I}),\ R(\rho)=e^{-\frac{m_{y}\omega_{\rho}}{2}\rho^{2}}\rho^{\tilde{l}}Hypergeometric_{1}F_{1}\left(-n_{\rho},\tilde{l}+2;m_{y}\omega_{\rho}\rho^{2}\right),\nonumber \\
E\left(l,n_{\rho},n_{z}\right) & =\omega_{\rho}\left(\tilde{l}+2n_{\rho}+2\right)=\sqrt{\frac{(l+1)^{2}}{6}+\frac{2}{15}N_{c}^{2}}+\frac{2\left(n_{\rho}+n_{z}\right)+2}{\sqrt{6}}.\tag{B-9}\label{eq:45}
\end{align}
Notice that $T^{(l)}(a_{I})$ is the function of the spherical part
which satisfies $\nabla_{S^{3}}^{2}T^{(l)}=-l(l+2)T^{(l)}$ since
$H_{y}$ could be rewritten with the radial coordinate $\rho$,

\begin{equation}
H_{y}=-\frac{1}{2m_{y}}\left[\frac{1}{\rho^{3}}\partial_{\rho}(\rho^{3}\partial_{\rho})+\frac{1}{\rho^{2}}\left(\nabla_{S^{3}}^{2}-2m_{y}\mathcal{Q}\right)\right]+\frac{1}{2}m_{y}\omega_{\rho}^{2}\rho^{2}.\tag{B-10}\label{eq:46}
\end{equation}
And we have used the quantum numbers $n_{z},n_{\rho},\tilde{l}$ to
denote the eigenvectors of the combined quantum system $H_{y}+H_{Z}$
as $|n_{z},n_{\rho},\tilde{l}\rangle$ in this paper.

\section*{Appendix C: Some calculations about the perturbed Hamiltonian}

The perturbed Hamiltonian (\ref{eq:17}) can be computed by (\ref{eq:2})
(\ref{eq:15}) or (\ref{eq:42}), equivalently, 

\begin{align}
-H_{Scalar}\left(t,X^{\alpha}\right)= & -\frac{1}{4}\left(2\pi\alpha^{\prime}\right)^{2}T_{8}\int_{\mathrm{D}8/\overline{\mathrm{D}8}}d^{4}xdzd\Omega_{4}e^{-\Phi}\sqrt{-\det g_{ab}}\mathrm{Tr}\left[g^{ac}g^{bd}\mathcal{F}_{ab}\mathcal{F}_{cd}\right]-\nonumber \\
 & \left\{ -\frac{1}{4}(2\pi\alpha')^{2}T_{8}\int_{\mathrm{D}8/\overline{\mathrm{D}8}}d^{4}xdzd\Omega_{4}e^{-\Phi^{(0)}}\sqrt{-\det g_{ab}^{(0)}}\mathrm{Tr}\left[g^{(0)ac}g^{(0)bd}\mathcal{F}_{ab}\mathcal{F}_{cd}\right]\right\} .\tag{C-1}\label{eq:47}
\end{align}
With the linear perturbation of gravity, we have,

\begin{equation}
g^{ab}=g^{(0)ab}-h^{ab},\ h^{ab}=g^{(0)ac}g^{(0)bd}h_{cd}.\tag{C-2}\label{eq:48}
\end{equation}
Therefore all the functions in (\ref{eq:47}) have been given in (\ref{eq:3})
(\ref{eq:14}) (\ref{eq:28}) (\ref{eq:38}). Rescale the formulas
in (\ref{eq:47}) as (\ref{eq:37}), we can obtained the following
result by direct computation,

\begin{equation}
H_{Scalar}\left(t,X^{\alpha}\right)=\frac{1}{4}\left(2\pi\alpha^{\prime}\right)^{2}T_{8}\int_{\mathrm{D}8/\overline{\mathrm{D}8}}d^{3}\boldsymbol{\mathrm{x}}d\boldsymbol{\mathrm{z}}d\Omega_{4}\mathcal{K}\left(t,\boldsymbol{\mathrm{x}},\boldsymbol{\mathrm{z}},X^{\alpha}\right),\tag{C-3}\label{eq:49}
\end{equation}
where

\begin{align}
\mathcal{K}\left(t,\boldsymbol{\mathrm{x}},\boldsymbol{\mathrm{z}},X^{\alpha}\right)= & C_{1}\mathrm{Tr}\bigg\{\frac{9Q^{2}\delta_{ij}\tau^{i}\tau^{j}}{8M_{E}^{2}}\left(2M_{E}^{2}-5\omega^{2}\right)G_{E}\left(t,\boldsymbol{\mathrm{x}}\right)+\frac{45k\omega Q\left(\hat{\boldsymbol{\mathrm{F}}}_{02}\tau^{1}-\hat{\boldsymbol{\mathrm{F}}}_{01}\tau^{2}+\hat{\boldsymbol{\mathrm{F}}}_{0z}\tau^{3}\right)}{8M_{E}^{2}}\nonumber \\
 & \times G_{E}\left(x\right)\lambda^{-1/2}+\bigg[-\frac{9}{64M_{E}^{2}}\bigg(\left(5\hat{\boldsymbol{\mathrm{F}}}_{0z}^{2}-5\hat{\boldsymbol{\mathrm{F}}}_{03}^{2}\right)k^{2}+\left(7\hat{\boldsymbol{\mathrm{F}}}_{0z}^{2}-3\hat{\boldsymbol{\mathrm{F}}}_{03}^{2}\right)M_{E}^{2}\nonumber \\
 & +\left(5\hat{\boldsymbol{\mathrm{F}}}_{03}^{2}+5\hat{\boldsymbol{\mathrm{F}}}_{0z}^{2}\right)\omega^{2}+\left(\hat{\boldsymbol{\mathrm{F}}}_{01}^{2}+\hat{\boldsymbol{\mathrm{F}}}_{02}^{2}\right)\left(5k^{2}-3M_{E}^{2}+5\omega^{2}\right)\bigg)G_{E}\left(x\right)\nonumber \\
 & -\frac{15i\omega\boldsymbol{\mathrm{z}}Q\hat{\boldsymbol{\mathrm{F}}}_{0i}\tau^{i}}{M_{E}^{2}}F_{E}\left(t,\boldsymbol{\mathrm{x}}\right)+\frac{3\boldsymbol{\mathrm{z}}^{2}Q^{2}}{32M_{E}^{2}}\bigg(40k^{2}\left(\left(\tau^{1}\right)^{2}+\left(\tau^{2}\right)^{2}-\left(\tau^{3}\right)^{2}\right)\nonumber \\
 & -\left(\left(\tau^{1}\right)^{2}+\left(\tau^{2}\right)^{2}+\left(\tau^{3}\right)^{2}\right)\times\left(6M_{E}^{4}-140\omega^{2}-M_{E}^{2}\left(16+15\omega^{2}\right)\right)\bigg)G_{E}\left(t,\boldsymbol{\mathrm{x}}\right)\bigg]\lambda^{-1}\nonumber \\
 & +\mathcal{O}\left(\lambda^{-3/2}\right)\bigg\}.\tag{C-6}\label{eq:50}
\end{align}
Notice that (\ref{eq:50}) is written in the unit of $U_{KK}=M_{KK}=1$,
so that $R^{3}=9/4$. The explicit formula of the glueball field $G_{E}\left(t,\boldsymbol{\mathrm{x}}\right)$
is needed in order to work out (\ref{eq:17}). The most simple way
is to solve its classical equation of motion from the action (\ref{eq:7}).
So we choose the real solution for $G_{E}\left(t,\boldsymbol{\mathrm{x}}\right)$
since it also appears in the perturbed metric (\ref{eq:9}) of the
bulk geometry, therefore we have

\begin{align}
G_{E}\left(t,\boldsymbol{\mathrm{x}}\right) & =\frac{e^{-ik_{\mu}x^{\mu}}+e^{ik_{\mu}x^{\mu}}}{2}=\cos\left(kx^{3}-\omega t\right)\nonumber \\
 & =\cos\left(\frac{k\boldsymbol{\mathrm{x}}^{3}}{\lambda^{1/2}}-\omega t\right),\tag{C-7}\label{eq:51}
\end{align}
so that the derivatives of $G_{E}\left(t,\boldsymbol{\mathrm{x}}\right)$
in (\ref{eq:3}) can be calculated as,

\begin{equation}
\partial_{\mu}G_{E}\left(t,\boldsymbol{\mathrm{x}}\right)=\frac{ik_{\mu}\left(e^{ik_{\mu}x^{\mu}}-e^{-ik_{\mu}x^{\mu}}\right)}{2}\equiv ik_{\mu}F_{E}\left(t,\boldsymbol{\mathrm{x}}\right),\tag{C-8}\label{eq:52}
\end{equation}
Notice that since the system is rotationally symmetric in $x^{i}$-space,
we have assumed that the momentum $k$ in (\ref{eq:51}) (\ref{eq:52})
has only one component along $x^{3}$ direction. In order to further
simplify (\ref{eq:50}), we calculate the following integrals appearing
in (\ref{eq:49}),

\begin{align}
\boldsymbol{\mathrm{I)}}\int_{-\infty}^{+\infty}d^{3}\boldsymbol{\mathrm{x}}d\boldsymbol{\mathrm{z}}Q\left(\boldsymbol{\mathrm{x}},\boldsymbol{\mathrm{z}}\right)^{2}G_{E}\left(t,\boldsymbol{\mathrm{x}}\right)= & \frac{1}{3}\frac{k^{2}\pi^{2}\rho^{2}}{\lambda}\mathrm{BesselK}\left[2,\frac{k\rho}{\lambda^{1/2}}\right]\cos\left(\omega t\right)\equiv\mathcal{I}_{1}\cos\left(\omega t\right)\nonumber \\
\simeq & \left[\frac{2}{3}\pi^{2}-\frac{1}{6}\frac{\pi^{2}\rho^{2}k^{2}}{\lambda}+\mathcal{O}\left(\lambda^{-2}\right)\right]\cos\left(\omega t\right),\tag{C-9}\label{eq:53}
\end{align}

\begin{align}
\boldsymbol{\mathrm{II)}} & \int_{-\infty}^{+\infty}d^{3}\boldsymbol{\mathrm{x}}d\boldsymbol{\mathrm{z}}\boldsymbol{\mathrm{z}}^{2}Q\left(\boldsymbol{\mathrm{x}},\boldsymbol{\mathrm{z}}\right)^{2}G_{E}\left(t,\boldsymbol{\mathrm{x}}\right)\nonumber \\
= & \frac{1}{9}\pi^{2}\rho^{2}\bigg\{\frac{k^{2}}{\lambda}\left(3Z^{2}+\rho^{2}\right)\mathrm{BesselK}\left[2,\frac{k\rho}{\lambda^{1/2}}\right]\nonumber \\
 & +2\mathrm{MeijerG}\left[\left\{ -\frac{1}{2},\#\right\} ;\left\{ 0,1\right\} ,\left\{ \frac{1}{2}\right\} ;\frac{k^{2}\rho^{2}}{4\lambda}\right]\bigg\}\cos\left(\omega t\right)\equiv\mathcal{I}_{2}\cos\left(\omega t\right)\nonumber \\
\simeq & \bigg[\frac{1}{3}\pi^{2}\left(2Z^{2}+\rho^{2}\right)+\frac{1}{36}\pi^{2}\rho^{2}\big(-6Z^{2}-5\rho^{2}+6\gamma\rho^{2}+6\rho^{2}\log\frac{k}{\lambda^{1/2}}+3\rho^{2}\log\frac{\rho^{2}}{4}\nonumber \\
 & -3\rho^{2}\mathrm{PolyGamma}\left[0,\frac{3}{2}\right]+3\rho^{2}\mathrm{PolyGamma}\left[0,\frac{5}{2}\right]\big)\frac{k^{2}}{\lambda}+\mathcal{O}\left(\lambda^{-2}\right)\bigg]\cos\left(\omega t\right),\tag{C-10}\label{eq:54}
\end{align}

\begin{align}
\boldsymbol{\mathrm{III)}} & \int_{-\infty}^{+\infty}d^{3}\boldsymbol{\mathrm{x}}d\boldsymbol{\mathrm{z}}\hat{\boldsymbol{\mathrm{F}}}_{03}^{2}G_{E}\left(t,\boldsymbol{\mathrm{x}}\right)\nonumber \\
= & \frac{1}{11520a^{2}\pi^{2}\rho^{2}}\bigg\{4\mathrm{MeijerG}\left[\left\{ -\frac{5}{2},\#\right\} ;\left\{ 0,1\right\} ,\left\{ \frac{1}{2}\right\} ;\frac{k^{2}\rho^{2}}{4\lambda}\right]\nonumber \\
 & +18\mathrm{MeijerG}\left[\left\{ -\frac{3}{2},\#\right\} ;\left\{ 0,2\right\} ,\left\{ \frac{1}{2}\right\} ;\frac{k^{2}\rho^{2}}{4\lambda}\right]\nonumber \\
 & +21\mathrm{MeijerG}\left[\left\{ -\frac{1}{2},\#\right\} ;\left\{ 0,3\right\} ,\left\{ \frac{1}{2}\right\} ;\frac{k^{2}\rho^{2}}{4\lambda}\right]\bigg\}\cos\left(\omega t\right)\equiv\mathcal{I}_{3}\cos\left(\omega t\right)\nonumber \\
\simeq & \bigg[\frac{1}{80a^{2}\pi^{2}\rho^{2}}+\frac{1}{8960a^{2}\pi^{2}}\bigg(-101+70\gamma+70\log\frac{k}{\lambda^{1/2}}+35\log\frac{\rho^{2}}{4}-35\mathrm{PolyGamma}\left[0,\frac{3}{2}\right]\nonumber \\
 & +35\mathrm{PolyGamma}\left[0,\frac{9}{2}\right]\bigg)\frac{k^{2}}{\lambda}+\mathcal{O}\left(\lambda^{-2}\right)\bigg]\cos\left(\omega t\right),\tag{C-11}\label{eq:55}
\end{align}

\begin{align}
\boldsymbol{\mathrm{IV)}} & \int_{-\infty}^{+\infty}d^{3}\boldsymbol{\mathrm{x}}d\boldsymbol{\mathrm{z}}\hat{\boldsymbol{\mathrm{F}}}_{01,02,0z}^{2}G_{E}\left(t,\boldsymbol{\mathrm{x}}\right)\nonumber \\
= & \frac{1}{11520a^{2}\pi^{2}\rho^{2}}\bigg\{13\frac{k^{3}\rho^{3}}{\lambda^{3/2}}\mathrm{BesselK}\left[3,\frac{k\rho}{\lambda^{1/2}}\right]+16\mathrm{MeijerG}\left[\left\{ -\frac{3}{2},\#\right\} ;\left\{ 0,1\right\} ,\left\{ \frac{1}{2}\right\} ;\frac{k^{2}\rho^{2}}{4\lambda}\right]\nonumber \\
 & +56\mathrm{MeijerG}\left[\left\{ -\frac{1}{2},\#\right\} ;\left\{ 0,2\right\} ,\left\{ \frac{1}{2}\right\} ;\frac{k^{2}\rho^{2}}{4\lambda}\right]\bigg\}\cos\left(\omega t\right)\equiv\mathcal{I}_{4}\cos\left(\omega t\right)\nonumber \\
\simeq & \bigg\{\frac{1}{80a^{2}\pi^{2}\rho^{2}}+\frac{1}{11520a^{2}\pi^{2}\rho^{2}}\bigg(-49+30\gamma+30\log\frac{k}{\lambda^{1/2}}+15\log\frac{\rho^{2}}{4}-15\mathrm{PolyGamma}\left[0,\frac{3}{2}\right]\nonumber \\
 & +15\mathrm{PolyGamma}\left[0,\frac{7}{2}\right]\bigg)\frac{k^{2}}{\lambda}+\mathcal{O}\left(\lambda^{-2}\right)\bigg\}\cos\left(\omega t\right),\tag{C-12}\label{eq:56}
\end{align}
where $\gamma$ is the Euler-Gamma constant. Using (\ref{eq:50}),
(\ref{eq:53}) - (\ref{eq:56}) and $\mathrm{Tr}\left\{ \tau^{i}\right\} =0$,
we can obtain,

\begin{align}
H_{Scalar}\left(t,X^{\alpha}\right)= & \mathcal{A}\kappa C_{1}\cos\left(\omega t\right)\bigg\{\frac{27}{4M_{E}^{2}}\left(2M_{E}^{2}-5\omega^{2}\right)\mathcal{I}_{1}\nonumber \\
 & +\bigg[\frac{120k^{2}-9\left(6M_{E}^{4}-140\omega^{2}-M_{E}^{2}\left(16+15\omega^{2}\right)\right)}{16M_{E}^{2}}\mathcal{I}_{2}+\frac{9\left(3M_{E}^{2}-5\omega^{2}+5k^{2}\right)}{32M_{E}^{2}}\mathcal{I}_{3}\nonumber \\
 & -\frac{9\left(M_{E}^{2}+15\omega^{2}+15k^{2}\right)}{32M_{E}^{2}}\mathcal{I}_{4}\bigg]\lambda^{-1}+\mathcal{O}\left(\lambda^{-2}\right)\bigg\}.\tag{C-13}\label{eq:57}
\end{align}
Inserting the expansion of large $\lambda$ of $\mathcal{I}_{1,2,3,4}$,
then (\ref{eq:17}) can be obtained.

\end{document}